\documentclass[%
 reprint,
 amsmath,amssymb,
 aps,
]{revtex4-1}

\usepackage{amsmath, amsthm, amssymb, amsfonts}
\usepackage{subfigure}
\usepackage{graphicx}
\usepackage{dcolumn}
\usepackage{bm}

\begin{document}

\preprint{}

\title{Entanglement of heterogeneous free fermion chains}

\author{Yuchi He}

\affiliation{Department of Physics, Carnegie Mellon University
 }%

\date{\today}

\begin{abstract}
We calculate the ground state entanglement entropy between two heterogeneous parts of a free fermion chain. The two parts could be XX chains with different parameters or an XX half chain connected with a quantum Ising  half chain.  It is shown that logarithmic behavior holds if the two parts are conformally critical.
In other cases, area law holds  with abundant subleading behaviors. In particular, when XX chain at Lifshitz point is connected with a conformally  or Lifshitz critical part, entanglement entropy converges algebraically with a fractional subleading index.  
\end{abstract}

\pacs{Valid PACS appear here}
\maketitle

\section{Introduction}
The importance of bipartite entanglement entropy (EE), as a concept or a tool, has been realized by quantum physics community in recent years. For illustration, EE is related to the Bekenstein~\cite{solodukhin2011entanglement} entropy of black holes and can be used to diagnose quantum criticality~\cite{vidal2003entanglement,pollmann2009theory}, topological orders~\cite{kitaev2006topological}, and localization transition~\cite{berkovits2012entanglement,li2016quantum, yu2016bimodal}.
In a quantum ergodicity assumption called ETH,  the same scaling behavior  shared by EE and thermal entropy bridges quantum mechanics and statistical mechanics~\cite{bhattacharya2013thermodynamical,palmai2014excited, lai2015entanglement}.  Among these studies, the scaling law of EE over partition's size usually plays a central role. It has been proved~\cite{hastings2007area} that the ground state EE of a 1-D gapped system is bounded, which is the foundation of a powerful numerical method named DMRG~\cite{white1992density,schollwock2011density}. Proofs or arguments of ground state area law for a gaped system in arbitrary dimension has also been presented ~\cite{srednicki1993entropy,eisert2010colloquium}. Solid results have also been obtained for 1-D  conformal critical systems~\cite{calabrese2009entanglement}, of which ground state EE diverges logarithmically with the length of the partition.

While the entanglement properties of homogeneous~\cite{vidal2003entanglement,pollmann2009theory,kitaev2006topological,bhattacharya2013thermodynamical,palmai2014excited, lai2015entanglement,hastings2007area,srednicki1993entropy,eisert2010colloquium, calabrese2009entanglement, gioev2006entanglement} and disordered systems~\cite{refael2004entanglement, lin2007entanglement, berkovits2012entanglement,li2016quantum, yu2016bimodal} are well understood,  the entanglement  properties of heterogeneous systems are not.   
Heterogeneous systems are of great interest in physics.
In thermodynamics, the contacting of two different systems is a common context. In quantum transport experiments,  the device is a heterogeneous structure.
The entanglement between heterogeneous subsystems gives a measure of the quantum fluctuations which link
them together and make observables near boundary  different from observables of bulk.  As will be made clear in this paper, the study of entanglement of heterogeneous systems may provide new perspectives on phase transition and quantum transport. 

\begin{figure}
\includegraphics[width=2 in] {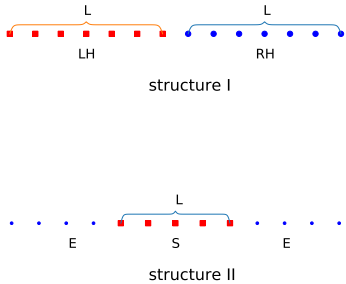}
\flushleft
\caption{Two kinds of structures to be studied. The enviroment (E) of structure II is infinitely large. Different symbols represent different parts of the chains. The chain is heterogeneous if the two parts are heterogeneous.}
\label{system}
\end{figure}

In this study, we are interested in the ground state entanglement of one-dimensional heterogeneous systems. 
For example, one may wonder what is the scaling law of  EE when two parts are critical and gapped respectively. Apart from the  properties of the two halves,  the conclusion may also depend on the types of interaction between the two parts.  A good starting point to study this topic is to look at a simple model. One choice is the heterogeneous system consisting of two XX  (lattice free fermion) chains with different potential and hopping coefficients. Another choice is connecting a quantum Ising (lattice BdG fermion) chain with an XX chain.  XX amd Ising chains can either be conformally critical, the corresponding CFTs are different. 
With the components chosen above, we study two kinds of heterogeneous structures (Fig.\ref{system}) in one dimension. One is that two parts are of the same size $L$, the other is that a finite subsystem of length $L$ is embedded in an infinite environment. For the systems described above, numerically exact calculations can be implemented for  $L$ large enough.  

Our main results are the properties of functions $S(L)$ (dependence of EE on $L$) in various situations.  Those situations can be classified by criticality of two parts of the chains.  XX chain with an external field can be conformal critical, Lifshitz critical and gapped.  Quantum Ising chain can be conformal critical and  topologically/trivially gapped.  Our results are as follows.

EE scales logarithmically if and only if the two parts are conformally critical.  Other situations follow area law. Within the area law,  the subleading behaviors of EE are abundant. While  subleading decay patterns of EE area law are always exponential~\cite{calabrese2010corrections} for homogeneous chains,  EE subleading terms of heterogeneous chains demonstrate either exponential or algebraic decay. The length scale of an exponential decay is found and argued to be the screening length.   The indices of algebraic decays are either integer or fractional. 

Fractional indices are found to be signatures of Lifshitz criticality. ( In contrast,  Lifshitz points of homogeneous free fermion chain have no non-trivial EE behavior~\cite{rodney2013scaling}.)  Remarkably, universal crossover of EE is observed when the two parts cross over from both gapped to both Lifshitz critical.
Besides heterogeneous chains with two Lifshitz critical parts,  fractional indices are also observed when a Lifshitz critical part meets with a conformal critical part. 

The outline of this paper is as follows.  Section \ref{s1} introduces the formulas to compute EE of free fermion (including BdG fermion)  from correlation functions. Section \ref{s2} mainly discusses EE of heterogeneous XX chains. In the beginning, the correlation functions are derived, and particle-hole symmetry of EE is discussed. After this preparation, EE  leading behaviors and  EE subleading oscillatory and  decay patterns are demonstrated and interpreted. EE leading and subleading behaviors are checked to be general for heterogeneous free fermion chains.  Lastly, section \ref{s3} investigates EE of XX/Ising heterogeneous chains.


\section{Entanglement of free fermions}\label{s1}

Bipartite entanglement entropy (EE) is defined as the von Neumann entropy of the reduced density matrix $\rho$: 
\begin{equation}
S=-\operatorname{Tr} (\rho \log \rho)
\end{equation}
Reduced density matrix $\rho$  is obtained by tracing the density matrix of the pure state over the degrees of freedom outside a given region. For one dimensional systems, there are two typical geometries (Fig.\ref{system}). The structure I is that the chain is finite with length $2L$ and the region with length $L$ is one of its halves. The structure II is that a finite subsystem of length $L$ is embedded in an infinite environment.  We are interested in how EE depends on the length $L$ of the region ($S(L)$). 

For lattice free fermion systems with charge conservation, EE can be calculated from equal time two-point Green function of either part~\cite{peschel2003calculation,cheong2004many}:  $C(r_{i},r_{j})=\langle c^{\dagger}_{r_{i}}c_{r_{j}}\rangle$ and
\begin{equation}\label{eq1}
S=\sum_{i}(-\lambda_{i} \log(\lambda_{i})-(1-\lambda_{i}) \log(1-\lambda_{i})) 
\end{equation}
where $\lambda_{i}$ are the eigenvalues of $C$.  

For BdG fermion, the expectation values of paired creation (annihilation) operators  can be non-zero. In this situation, the two-point Green functions $F(r_{i},r_{j})=\langle c^{\dagger}_{r_{i}}c^{\dagger}_{r_{j}}\rangle$ are also needed. Let $\nu_{i}$ be the eigenvalues of the matrix $(2C+2F-1)(2C-2F-1)$ and the EE is given by Eq.~(\ref{eq1}) with $\lambda_{i}=\frac{1+\sqrt{\nu_{i}}}{2}$. 

Using formulas above,  the ground state EE scaling laws of homogeneous free fermion chains have been studied~\cite{vidal2003entanglement}. The key result is that EE diverges with $L$ if and only if the system is conformal critical with leading term scales logarithmically with $L$.  The logarithmic behavior can also be derived from conformal field theory and generalized to generic 1-D local Hamiltonians~\cite{calabrese2009entanglement}.  For structure II, conformal field theory predicts that the critical degree of freedom contributes  $\frac{c}{3} \log L$ to the EE leading term, where $c$ is the central charge of the CFT's Virasoro algebra.  (For a given Hamiltonian on  structure I, the EE leading term  is always half of that on structure II.)   Applying the conclusion to a homogeneous free fermion chain with charge conservation,  the corresponding EE leading term on structure II  is  $\frac{m}{3} \log L$, where $m$ is the number of  conformal critical degrees of freedom and equals the number of pairs of Fermi surfaces (points). Near each pair of Fermi surfaces, there is an effective  free boson CFT with central charge $c=1$. $m$ can be changed by tuning chemical potential and the transition points are called  Lifshitz critical points.  
For a Lifshitz point separating $m=n$ and $m=n-1$ phase,  EE leading term right at the transition (Lifshitz) point is trivially same as the side with $m=n-1$~\cite{rodney2013scaling}.
As another illustration of ground state EE scaling law,  the EE leading term of critical Kitaev BdG fermion chain~\cite{kitaev2001unpaired} is $\frac{1}{6} \log L$ on structure II.  The critical phase is described by free Majorana fermion (Ising) CFT, where $c=1/2$. 

In the following, we apply Eq.~\ref{eq1} to study ground state entanglement of heterogeneous free fermion chains.

\section{Entanglement of of heterogeneous XX chain}\label{s2}

\subsection{Hamiltonian, single-particle eigenstates and equal time correlation matrix}
In section \ref{s2}, we mainly deal with  nearest-neighbor hopping free fermion chain: 
\begin{equation} 
H=\sum_{i} (-t_{i}c^{\dagger}_{i}c_{i+1}+h.c.+h_{i} c^{\dagger}_{i}c_{i}) 
\end{equation}

The chain is heterogeneous in the sense that $h$ and $t$ are homogeneous in two parts  respectively but differ from one part to the other.  At their mutual boundary, the two parts are linked by hopping with parameters $t_{interface}$. In what follows, we keep $t$ to be the same everywhere  for structure II while leaving it possibly different among left part, right part and interface of structure I.  We solve this Hamiltonian for both structures ~(Fig.\ref{system}) and study the ground state entanglement entropy between the two parts.  

The full chain can be characterized by the criticality of its parts.  A part is denoted as ``CFT" if there are Fermi surfaces.  If the Fermi energy is right at the band bottom or top,  the part is denoted as ``Lifshitz".  Otherwise, it is denoted as ``gapped".

For the heterogeneous system, the single-particle states are piecewise functions of the two parts. In each part, they are  linear combinations of the following basis: $e^{-\beta r}$, $e^{i k r}$ or $(-1)^r e^{-\beta r}$. The linear combinations and the values of $k$ and $\beta$ are determined by matching energies and boundary conditions. For example, single-particle states of structure II are listed in Fig.\ref{ESband}. The quantities relevant to behaviors of EE are $k$ and (or) $\beta$ of the single-particle states with Fermi energy. Except for trivially fully filled or empty systems,  a CFT part is characterized by Fermi vector $k_{F}$ and a gapped  part is characterized by inverse Fermi deay length $\beta_{F}$. A Lifshitz part is denoted by "0" or $\pi$. We refer a system on structure I as $(a_{LH}, b_{RH})$ or $A_{LH}/B_{RH}$. $a_{LH}$ and $b_{RH}$ are $k_{F}$ or $\overline{\beta_{F}}$ of left and right parts. The overline in $\overline{\beta_{F}}$  is used to distinguish $\beta_{F}$ from $k_{F}$. For example, $(\overline{0.223},\frac{2\pi}{3})$ means the right chain has  Fermi vector $|k_{F}|=\frac{2\pi}{3}$, while the left is gapped with $\beta_{F}=0.223$.  $A_{LH}$ and $B_{RH}$ can be ``CFT", ``Lifshitz" or ``gapped". Similarly, $(a_{E}, b_{S}, a_{E})$  and $A_{E}/B_{S}/A_{E}$ are used to denote  systems on structures II. $A_{E}$ and $a_{E}$ denote environments while $b_{S}$ and $B_{S}$ denote subsystems.

\begin{figure}
\includegraphics[width=2 in] {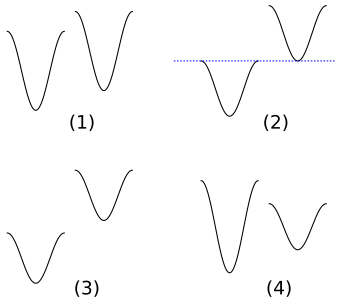}
\flushleft
\caption{Possible configurations. Two parts of each figure are band structures of the two parts of the system respectively. The two band structures may either describe LH, RH  for structure I or S, E for structure II.  One may also change the sign of $t$ to flip the bands. Fixing the Fermi energies of those configurations determines the state of the systems. In the second sub-figure,  when the Fermi surface is right at the dashed line, the two systems are both at Lifshitz point.  }
\label{possibleconfigurations}
\end{figure}

Equal time correlation matrices can be constructed  from single-particle eigenstates: $C(r_{i},r_{j})=\sum_{E(m)<E_{F}}\psi_{m}(r_{i})\psi_{m}(r_{j})$. The $C$ matrix is a summation of a Toeplitz matrix  and a Hankel matrix: $C(r_{i},r_{j})=D(r_{i}-r_{j})+D(r_{i}+r_{j})$.  Depending on types of occupied single particle states, $D(r)=\sum_{k} f(k) \cos(k r)$ or $D(r)=\sum_{\beta} g(\beta) \cosh(\beta r)$ or $D(r)=\sum_{k} f(k) \cos(k r)+ \sum_{\beta}g(\beta) \cosh(\beta r)$. $f(k)$ and $g(k)$  depend on the details of the model.  For scattering states in structure II, the summation is replaced by integration.  $k_F$ and (or) $\beta_{F}$ enter the summation (integration) as limits.

The EE is invariant under particle-hole transformation  $h_{i}=-h_{i}$. This symmetry is clear in the spin 1/2 representation by Jordan-Wigner transform.  This symmetry of ground state EE can be used to reduce the number of  situations to be calculated.  Also, it is helpful for determining the EE's oscillatory period.

\begin{figure}
\includegraphics[width=2 in] {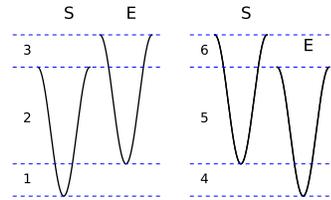}
\flushleft
\caption{Classification of the single-particle states of structure II.  Single-particle states exist in the energy range exapnded by the  band of subsystem and the band of environment (Fig.~\ref{possibleconfigurations}). The unnormalized single-particle wave functions of each region  are as follows. 1:$(PW, e^{-\beta|x|})$; 2, 5:$(PW,PW)$; 3 :$((-1)^{x} \cosh(\beta x) \ or\ (-1)^{x} \sinh(\beta x), PW)$;  4 :$(\cosh(\beta x) \ or\ \sinh(\beta x), PW)$;  6: $(PW,(-1)^x e^{-\beta|x|})$. The first components denote subsystem. PW means plane waves. Due to the potential flipping symmetry of EE and for computational convenience, we calculate the situations when the Fermi surface are in the region 1,  4, 5.}
\label{ESband} 
\end{figure}

\subsection{Entanglement entropy: leading terms and subleading oscillatory pattern}
\begin{figure*}
\includegraphics[width=4.5 in] {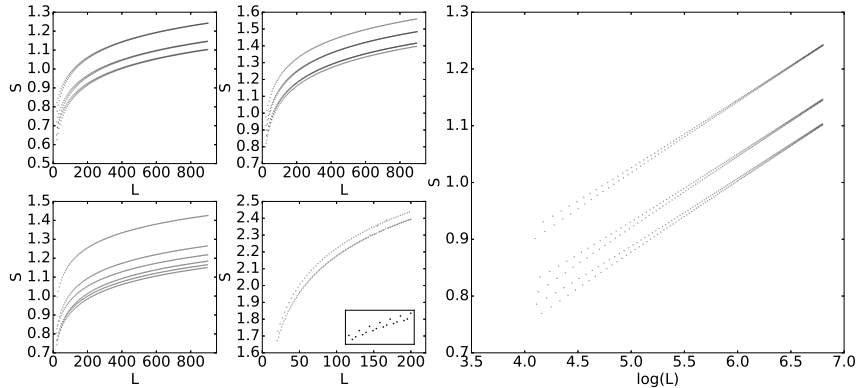}
\flushleft
\caption{ $S(L)$ of $(CFT/)CFT/CFT$ systems. Two small figures of the first row correspond to $(\frac{5\pi}{6},\frac{\pi}{2})$, $(\frac{2\pi}{3},\frac{\pi}{2})$  with uniform $t$. Their periods are both 6, but the numbers of branches differ.  
The first small figure of the second row shows that tuning $t_{RH}/t_{LH}$ away from 1 ($t_{RH}/t_{LH}=1.3512$) breaks the degeneracy of $(\frac{5\pi}{6},\frac{\pi}{2})$. The second shows the case $(\frac{\pi}{2},\frac{2\pi}{3},\frac{\pi}{2})$ with uniform $t$. Its inset is a detailed look at the lower branch, which shows the period is indeed 3. The large figure is $\log L\ vs.\ S$ fitting of the $(\frac{5\pi}{6},\frac{\pi}{2})$ figure. }
\label{doublegapless}
\end{figure*}

By numerical calculation, we obtain entanglement entropy of  structure I and structure II for different partition size $L$.  We find that EE  leading term of structure I is half of that of structure II if the bulks and the interface(s) are respectively the same.   EE diverges if and only if the two parts are conformally critical (Fig.~\ref{doublegapless}). The corresponding leading term is $\frac{c^{\prime}}{3}\log(L)$, where $c^{\prime}$ is determined by the transmission ratio of the mode at the Fermi point~\cite{eisler2010entanglement}.  Otherwise, the leading  behavior is area (constant) law.  The subleading term of logarithmic law is a possibly oscillatory o(1) term. The subleading term of area law may also be modulated by oscillation.  

The subleading term is oscillatory only if either part of structure I or the subsystem of structure II is CFT, due to the existence of Fermi vectors.  In contrast, for an infinite homogeneous free fermion chain, EE subleading term has no oscillatory behavior~\cite{susstrunk2013free}. The period of the oscillation is found to be determined by $k_{F,L}$, $k_{F,R}$ (structure I) and $k_{F,S}$ (structure II), while  
$k_{F,E}$ (structure II) is not related to the oscillation.

The oscillation pattern is given by the wave vector(s) $2k_{F,j}$, where $j$ denotes LH, RH or S.  The factor 2 comes from the fact that EE is invariant under particle-hole  transformation $|k_{F}| \rightarrow \pi-|k_{F}|$. 
The oscillation is strictly periodic if and only if $\frac{\pi}{k_{F,j}}$ is (are) rational and the period is (are) given by $T_{j}$, which is the denominator of the  irreducible fraction $\frac{\pi}{k_{F,j}}$.      For $CFT/CFT$ systems,  the period is the least common multiple of  $T_{LH}$ and  $T_{RH}$ (Fig.\ref{doublegapless}).  Note that $k_{F,L}$ and $k_{F,R}$  influence the oscillatory pattern symmetrically since one can either pick reduced density matrix of  the left or right part  to compute the bipartite EE. (Fig.\ref{doublegapless}).  Once the oscillation is periodic, the array $S(L)$ divides into several non-oscillatory branches. For $(CFT/)CFT/CFT$ systems, the number of branches may be smaller than the period. However, such degeneracy is not robust under tuning parameters $t_{i}$ in the two parts or at the interface (Fig.\ref{doublegapless}). 

We note that choosing oscillation to be strictly periodic is convenient for studying EE subleading terms' decay pattern.  

\subsection{Entanglement entropy: subleading decay pattern of area law}\label{key}
With the exception of $(CFT/)CFT/CFT$ systems,  the numerical results indicate that $S(L)$ converges to a single limit.  The patterns of convergence are observed to be either exponential or algebraic decays. 

\begin{figure*}
\includegraphics[width=4.5 in] {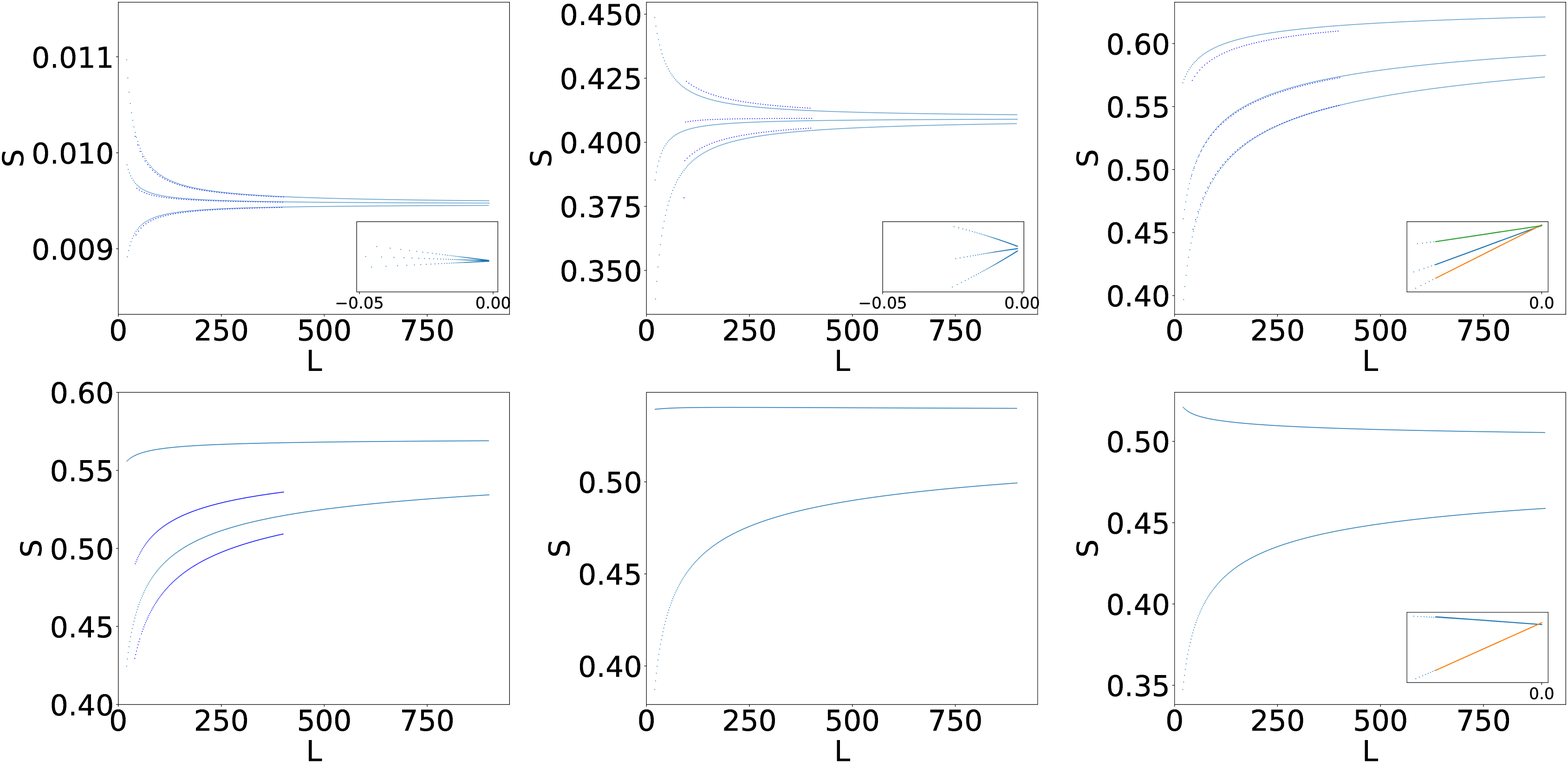}
\flushleft
\caption{$(gapped/)CFT/gapped$ and $(Lifshitz/)CFT/Lifshitz$ systems on structure I and structure II.  The EE of the structure II (blue curves) is divided by 2 to be compared with structure I. The first row shows transition from  $(gapped/)CFT/gapped$ to $(Lifshitz/)CFT/Lifshitz$. $t$ is uniform. The systems from left to right are $((\overline{2.559},)\frac{\pi}{3},\overline{2.559})$, $((\overline{0.3149},)\frac{\pi}{3},\overline{0.3149})$,  $((\pi,)\frac{\pi}{3},\pi)$. 
 When we are tuning $\beta_{F}$, a transition of monotonicity of the middle branch happens.  As the gap vanishes ($\beta_{F}\rightarrow 0$), the asymptotic behavior changes.
 The second row shows that tuning $t_{LH}$ changes the monotonicity of one branch of  $(\frac{\pi}{2},0)$ and makes it well fitted by index $\frac{1}{3}$. $t_{LH}$ of the three figures are $1$, $1.22$ and $1.5$. The interface $t$ does not influence the  shape of branches. 
 Insets without solid lines are $-L^{-1}\ vs.\ S  $ while those with solid lines are $-L^{-1/3}\ vs.\ S$ fitting. The solid lines are the linear regression of results from structure I. The lines in each inset cross at 0.}
\label{m1singlemarginal}
\end{figure*}

For $(gapped/)CFT/gapped$ systems,  we find the convergence of each  $S(L)$ branch is algebraic (Fig.\ref{m1singlemarginal}): 
\begin{equation}\label{integeralgebraic}
S=S_{0}+\frac{d}{L},
\end{equation}
where $d$ is different for different branches.  Tuning the hopping parameters or gap can change the sign $d$ of some branch (Fig.\ref{m1singlemarginal}).

\begin{figure}
\includegraphics[width=3.5 in] {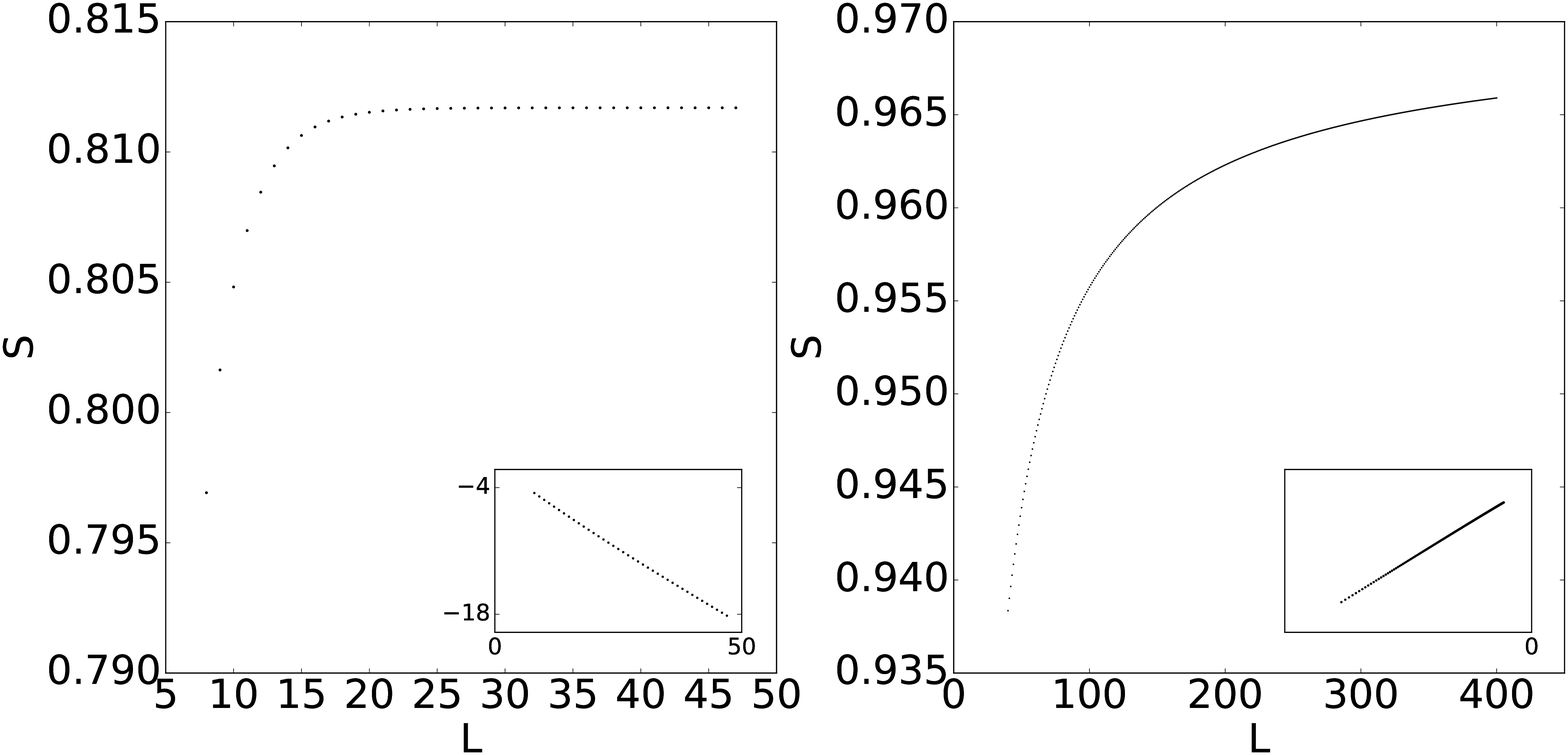}
\flushleft
\caption{$CFT/gapped/CFT$ and $CFT/Lifshitz/CFT$ systems of structure II. The first figure represents a gapped subsystem $(\frac{2\pi}{3},\overline{0.223},\frac{2\pi}{3})$.  Inset is $L\ vs.\ \log(S-S_{0})$ fitting. Slope is fitted to be $0.346$, whereas our estimation is $0.446$.  The second figure is: $(\frac{2\pi}{3},0,\frac{2\pi}{3}$.). Inset is $-L^{-5/6}\ vs.\ S$ fitting. }
\label{m2eglsinglemarginal}
\end{figure}

For $CFT/gapped/CFT$ systems, $S(L)$  is not oscillatory and $S(L)$ shows an exponential convergence (Fig.\ref{m2eglsinglemarginal}):
\begin{equation}\label{exponential}
 S=S_{0}-d^{\prime}e^{-\alpha L},
\end{equation} 
where $d^{\prime}$ is a constant. Here, $\alpha \approx 2 \beta_{F,S}=2\operatorname{arccosh}(1+\frac{g}{2t_{S}})$, where $g$ is the gap and $t_{S}$ is the hopping amplitude. The inverse decay length $2 \beta_{F,S}$ can be considered as analytical continuation of oscillation wave vector $2 k_{F,S}$ in the previous subsection.

\begin{figure}
\includegraphics[width=3.5 in] {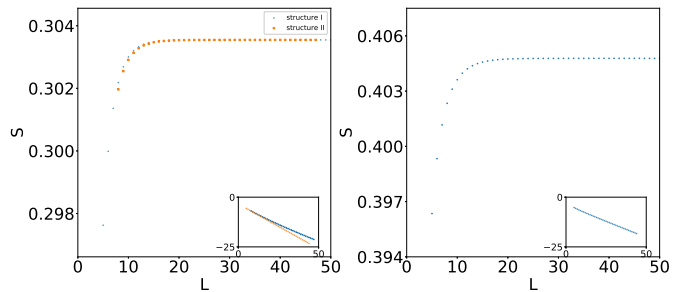}
\flushleft
\caption{$S(L)$ of $gapped/gapped/gapped$ systems. The gap of the figures are all set to be $0.05$. The first figure describes structure I and structure II with uniform $t=1$. EE of structure II is divided by 2 to compare with structure I. The second figure is for structure I, tuning $t_{RH}=2$. Insets are  $L\ vs.\ \log(S-S_{0})$ fitting. The fitting slopes are $0.449$, $0.363$ for structure I and structure II in figure 1 and $0.331$ in figure 2, comparing to our estimation $0.446$,$0.446$ and $0.316$.}
\label{doublegapped}
\end{figure}

For $(gapped/)gapped/gapped$ systems,   $S(L)$  also converges exponentially (Fig.\ref{doublegapped}). Similarly to the previous situation, $\alpha$ is estimated as $2 \beta_{F,S}=2 \operatorname{arccosh}(1+\frac{g}{2t_{S}})$ for structure II.  For structure I, $\alpha$  is approximately  $2 \min (\beta_{F,LH},\beta_{F,RH})= 2\operatorname{arccosh}(1+\frac{g}{2\max(t_{LH},t_{RH})})$. 

Remarkably, we find that systems with Lifshitz part(s)  have algebraic  EE subleading behavior with  fractional indices: 
\begin{equation}
S(L)=S_{0}-\frac{d^{\prime \prime}}{L^{\gamma}}
\end{equation}
Systems with Lifshitz part(s) are classified and  labeled as :

[a] $CFT/Lifshitz/CFT$ 

[b] $(Lifshitz/)CFT/Lifshitz$

[c] $(Lifshitz/)Lifshitz/Lifshitz$. (Fig.~\ref{possibleconfigurations}. (2)) 

The $S(L)$ of [a], [c] has a single branch while $S(L)$ of [b] has multiple branches.  We extract $\gamma$ by fitting $\frac{dS(L)}{dL}$ of each branch in a log-log plot. The fitting results are validated by the linearity of $-1/L^{\gamma}\ vs.\ S$ plots. In such plots, all the branches of [b] cross at one point when $-1/L^{\gamma}$ is extrapolated to zero.

\begin{figure}
\includegraphics[width=2.5 in] {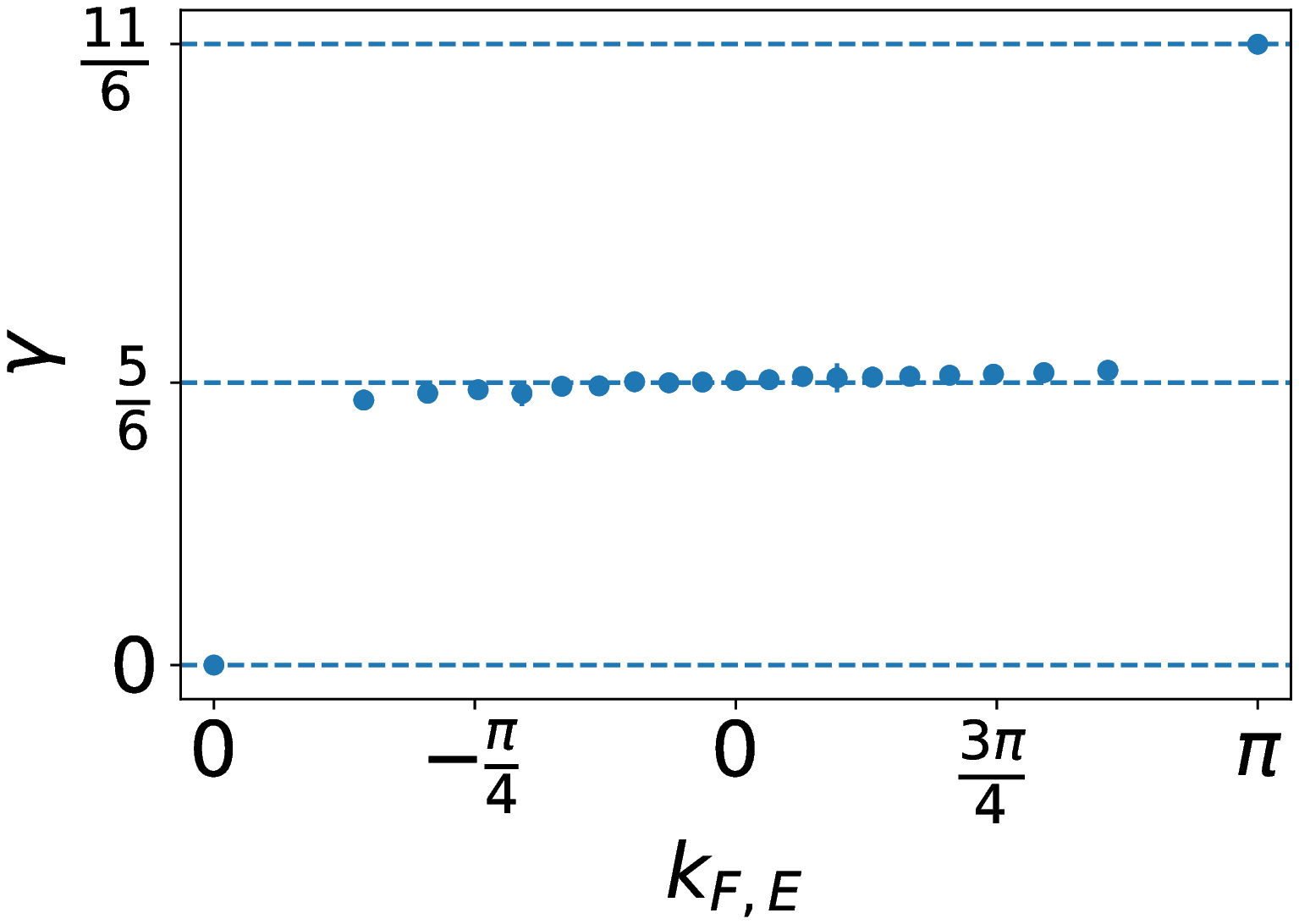}
\flushleft
\caption{Subleading indices fitting  of  $CFT/Lifshitz/CFT$ and $Lifshitz/Lifshitz/Lifshitz$ systems. All the systems can be labeled as $(k_{F,E},0,k_{F,E})$. Two ends are the $Lifshitz/Lifshitz$ limit. The left end is trivially direct product states. The EE curves of the right end is showed in Fig  \ref{doublemarginal}.} 
\label{indextransition}
\end{figure}

\begin{figure}
\includegraphics[width=2.5 in] {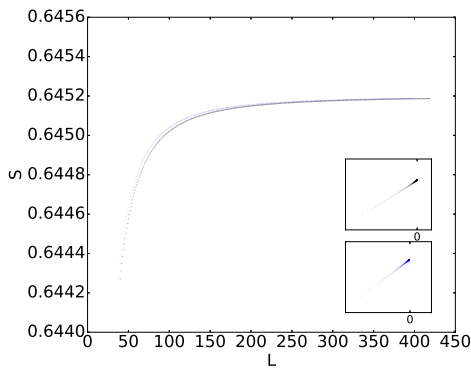}
\flushleft
\caption{$S(L)$ of  $(Lifshitz/)Lifshitz/Lifshitz$ systems. The EE of the structure I is multiplied by 2 to be compared with structure II. The blue curve (slightly upper) is for structure II. The insets are $-L^{-11/6}\ vs.\ S$ fitting. }
\label{doublemarginal}
\end{figure}

The fitted $\gamma$ for [a], [b] and [c] are close to rational number  $\frac{5}{6}$,  $\frac{1}{3}$ and $\frac{11}{6}$ respectively. Errors of the fitting are estimated and listed in Table~\ref{table}. We demonstrate the $-1/L^{\gamma}\ vs. \ S$ plots in Figs.~\ref{m1singlemarginal}, ~\ref{m2eglsinglemarginal} and~\ref{doublemarginal}. The fitting results of [a] and [c] on structure II is shown in Fig.\ref{indextransition}.  For [b], like $d$ in Eq.~\ref{integeralgebraic},  $d^{\prime\prime}$ of some branch can also change sign by tuning hopping parameters.  When $d$ and $d^{\prime\prime}$ is close to 0, the branch appears flat (Fig.~\ref{m1singlemarginal}), and gives an indetermined value for $\gamma$. 

To have some understanding of the fractional subleading behavior,  it is helpful to look at crossover from gapped to Lifshitz phase.  The above [a], [b] and [c] systems can be approached by decreasing the gap of  $[\text{a}^{\prime}]$,$[\text{b}^{\prime}]$ and $[\text{c}^{\prime}]$  respectively, where

$[\text{a}^{\prime}]$ $CFT/gapped/CFT$

$[\text{b}^{\prime}]$ $(gapped/)CFT/gapped$ 

$[\text{c}^{\prime}]$ $(gapped/)gapped/gapped$  

During this procedure,  the numbers of electrons (or holes) in the gapped part and the particle number fluctuations increase. The EE  is a measure of fluctuation~\cite{song2010general} and is expected to increase with more fluctuation.  As the gap of one part vanishes,  the part is Lifshitz critical and those almost extended bound states $\beta  \approx   0$  are occupied.  ($1/\beta_{F}$ is the length scale of the interface, where the gapped part has some electrons or holes.)  Intuitively, such almost extended states make  EE more sensitive to $L$  in the large $L$ limit. Hence, the convergence is expected to be slower and the subleading behavior (Eq.~\ref{integeralgebraic} and Eq.~\ref{exponential}) should change (Fig.\ref{m1singlemarginal}, Fig.\ref{m2eglsinglemarginal}). For the crossover from $[\text{a}^{\prime}]$ to [a] and  from $[\text{b}^{\prime}]$ to [b],  the change is from algebraic decay with  index 1 to algebraic decay with an index smaller than 1. The fitting results are $\gamma=\frac{5}{6}<1$ for [a], and  $\gamma=\frac{1}{3}<1$ for [b]. For the crossover from $[\text{c}^{\prime}]$ to [c], the change is from exponential decay to algebraic decay with  $\gamma=\frac{11}{6}$.  

\begin{figure}
\includegraphics[width=2.2 in] {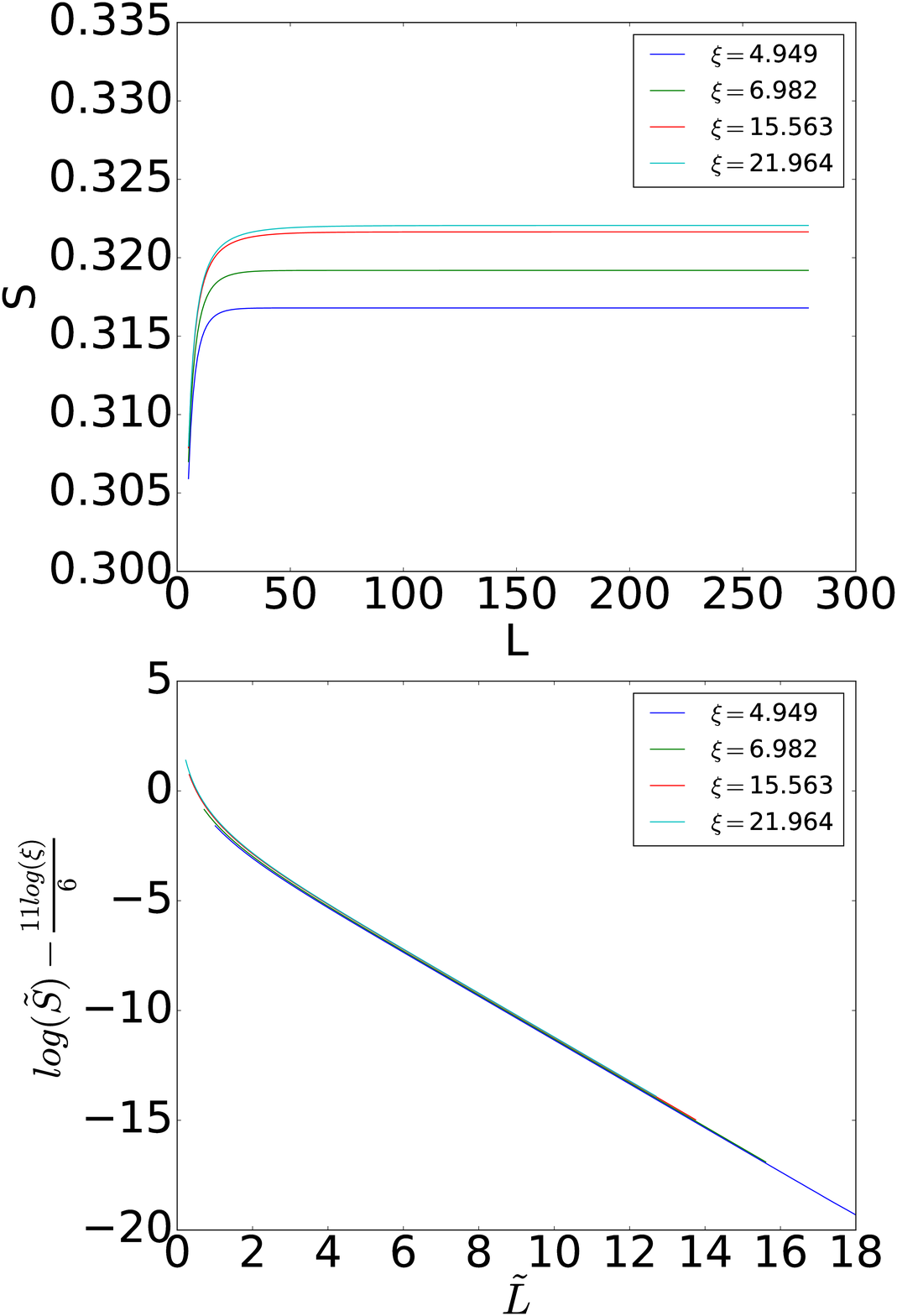}
\flushleft
\caption{Universal crossover of  $gapped/gapped$ systems (structure I). For $L<\xi$, the system behaves as  $Lifshitz/Lifshitz$ system. The second figure shows the scaling behavior: $\tilde{S}=\xi^{11/6} f(\tilde{L})$, where $\tilde{S}=S(\infty)-S$, $\tilde{L}=L/\xi$.}
\label{doublegappedcrossover}
\end{figure}

The above crossovers feature the divergence of length scale.  Universal crossover is expected to be observed.  For a given  $(gapped/)gapped/gapped$ systems with a small $\beta_{F}$ ,  the subleading term of $S(L)$ seems to be algebraic as $(Lifshitz/) Lifshitz/Lifshitz$ systems for small $L$,  but really is exponential for large $L$. $S(L)$ is a universal function after being rescaled by the length scale $1/\alpha$ of each $gapped/gapped/gapped$ system (Fig.~\ref{doublegappedcrossover}).  The perfect collapse of different curves implies $11/6$ is a very good fitting.  

The subleading behaviors and the values of algebraic decay indices are believed to be general for heterogeneous free fermion systems.  We check this by adding next-nearest-neighbor hopping and  considering two bands (SSH) model on structure I. 

\section{Entanglement of heterogeneous quantum Ising/XX chain}\label{s3}
In this section, we calculate EE for one kind of heterogeneous quantum Ising/XX chain which can be mapped to BdG fermion.  We study this model on structure I.
The Ising part  and the XX part of Hamiltonian are respectively: 
\begin{equation} 
\begin{split}
&H_{QIsing}= \sum_{i} (-\sigma_{x,i}\sigma_{x,i+1}+g\sigma_{z, i})\\
&H_{XX}=\sum_{i} (-2t (\sigma_{x, i}\sigma_{x, i+1}+ \sigma_{y, i}\sigma_{y, i+1})+h\sigma_{z,i})
\end{split}   
\end{equation}

The coupling between two parts is an XX coupling with coefficient $t_{interface}$. It is found that that the amplitude of $t_{interface}$ and switching to Ising type coupling do not influence the results qualitatively. 

\begin{figure*}
\includegraphics[width=4.5 in] {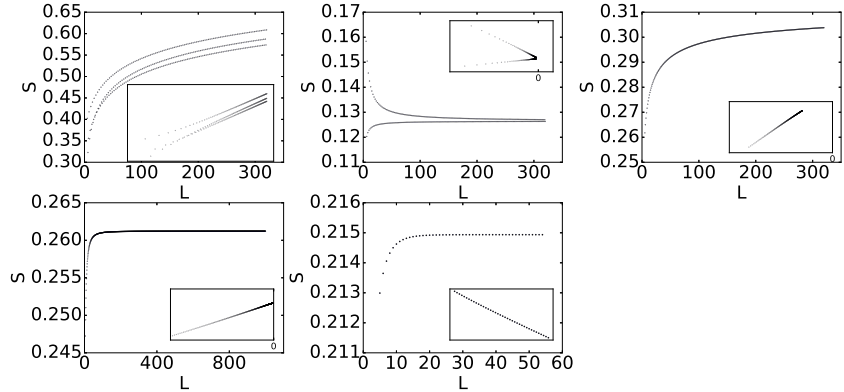}
\flushleft
\caption{$S(L)$ of  XX/Ising systems. From left to right, the figures of first row are $CFT_{XX}/CFT_{Ising}$, $CFT_{XX}/gapped_{Ising}$ and $Lifshitz_{XX}/CFT_{Ising}$ while the figures of  second row are $gapped_{XX}/CFT_{Ising}$ and $gapped_{XX}/gapped_{Ising}$. By the same order, the insets are are $\log L\ vs.\ S$, $-L^{-1}\ vs.\ S  $,  $-L^{-1/3}\ vs.\ S$, $-L^{-2}\ vs.\ S$ and  $L\ vs.\ \log(S-S_{0})$ fitting. }
\label{xxising}
\end{figure*}

The results (Fig.~\ref{xxising}) are quite similar to the last section. The oscillatory period is determined by the Fermi vector of the XX part. The EE diverges (logarithmically) if and only if both parts are conformally critical. Subleading terms are found to be either exponential and algebraic decay. Algebraic decay with a fractional index is again a signature of Lifshitz criticality. We only observe integer index 1 in the last section, but for Ising/XX chain, we observe both 1 and 2. It is found that whether the gapped Ising phase is topological or not is irrelevant to the behavior of  EE.  

\section{Summary and Discussion}

In summary, we have investigated the EE scaling law of heterogeneous free fermion chains.  The EE  behaviors of various situations are summarized in the Table~\ref{table}. We find that Logarithmic law applies for $(CFT/)CFT/CFT$ systems while area law applies for other situations. Remarkably, we also find both exponential and algebraic area law subleading behaviors.   The algebraic indices are found to be non-integer when one of the subsystems is Lifshitz critical.  Algebraic behavior is consequence of divergence of length scale; to see this, universal crossover near Lifshitz criticality is illustrated.  
We noted  that a special case of our structure I has been studied~\cite{eisler2015entanglement}. Their focus on the effective central charges of logarithmic behavior together with our results of subleading EE behaviors complementarily present the features of heterogeneous free fermion systems.

\begin{table}

 \centering
 \subtable[XX chain, structure I]{
       \begin{tabular}{|l|l|l|}
       \hline
LH& RH& EE behavior\\
 \hline
CFT & CFT & logarithmic \\
 \hline
CFT & Lifshitz & $\gamma=\frac{1}{3}\pm0.04$ \\
 \hline
CFT & gapped & $\gamma=1\pm0.02$ \\
 \hline
Lifshitz &Lifshitz & $\gamma=\frac{11}{6}\pm0.005$ \\
 \hline
gapped&gapped & exponential \\
 \hline
\end{tabular}
        \label{tab:firsttable}
 }
 \qquad
 \subtable[XX chain, structure II]{        
        \begin{tabular}{|l|l|l|}
       \hline
S& E& EE behavior\\
 \hline
CFT & CFT & logarithmic \\
 \hline
CFT & Lifshitz & $\gamma=\frac{1}{3}\pm0.02$  \\
 \hline
CFT & gapped & exponential \\
 \hline
Lifshitz &CFT& $\gamma=\frac{5}{6}\pm0.01$ \\
 \hline
Lifshitz &Lifshitz & $\gamma=\frac{11}{6}$ \\
 \hline
gapped&CFT & $\gamma=1\pm0.03$ \\
 \hline
gapped&gapped & exponential \\
 \hline
\end{tabular}
        \label{tab:secondtable}
 }
 \qquad
 \subtable[XX/quantum Ising chain]{        
        \begin{tabular}{|l|l|l|}
       \hline
XX & QI & EE behavior\\
 \hline
CFT & CFT & logarithmic \\
 \hline
CFT & gapped & $\gamma=1\pm0.04$ \\
 \hline
Lifshitz &CFT & $\gamma=\frac{1}{3}\pm0.04$ \\
 \hline
gapped & CFT & $\gamma=2\pm0.01$ \\
 \hline
gapped&gapped & exponential \\
 \hline
\end{tabular}
        \label{tab:secondtable}
 }
\caption{Summarization of leading and subleading EE behaviors. The errors of algebraic decay indices are formally written as $\pm \sigma$. $\sigma$ are estimated from system ensembles: $\sigma=\sqrt{\sum_{i}(\gamma_{i}-\gamma)^2/N}$.  Each ensemble includes more than 10 points. Those points are away from crossover region and (or) accidental flat region. The indices of (a) and (c) are fitted from $S(L)$ between $L=300$ and $L=500$.
The indices of (b) are fitted from $S(L)$ between $L=200$ and $L=400$.  }
\label{table}
 \end{table}

The heterogeneous systems entanglement problem can be considered as a further generalization of the entanglement  problem of homogeneous systems with localized impurity~\cite{peschel2005entanglement,eggert1992magnetic, zhao2006critical, sorensen2007quantum, igloi2009entanglement, saleur2013entanglement}. 
Here, we solve the heterogeneous problem for free fermion systems. One open question is to get the fully analytical result by finding or conjecturing spectral function in Eq.~\ref{eq1} ~\cite{susstrunk2013free, ossipov2014entanglement}. We conjecture that those fractional exponents might be found to be exact fractions. It is also interesting to consider the problem for interacting systems.  In particular, some interaction might  change the fixed point.  For example, the algebraic decay is "gapped out" to become exponential decay and the junction becomes  open boundary in RG sense.  There is an even more interesting possibility, that interaction drives the interface to another "algebraic" fixed point.  In this situation, it's interesting to seek a field theoretical derivation formalism.  Conformal field theory and conformal perturbation theory have been used to  explain  EE behaviors of  homogeneous systems ~\cite{calabrese2009entanglement} and homogeneous systems with localized impurity ~\cite{eggert1992magnetic}.  However,  because  Lifshitz criticality rather than conformal criticality plays the main role in the current problem,  formalism beyond CFT is possibly needed. 

The heterogeneous systems can also be considered as intermediate between homogeneous systems and disordered systems.  Thus, the study of EE of heterogeneous systems might be able to shine light on the problem of many-body localization.  One might use EE and other mutual information to decide if two parts of  heterogeneous systems  are well connected.  Likewise, in resonant cluster picture of many-body localization~\cite{potter2015universal}, there might be ways to use entanglement structure to define if two subsystems belong to the same  cluster.   The entanglement and quantum fluctuation involving dynamic exponent $\neq1$ is also interesting. It might be studied in systems such as multilayer graphene~\cite{mccann2013electronic}.

\begin{acknowledgments}
 The author is grateful to Prof. Roger Mong for helpful discussions through the project and advice on the revision of this paper, Zhi Li for helpful discussion, Prof. David Pekker for motivating his interest in the study of EE of nonuniform systems. The author also acknowledges Prof. Ingo Peschel for interest and  discussion. 
\qquad

\qquad

\qquad

\qquad

\qquad

\qquad

\qquad

\qquad

\qquad

\qquad

\end{acknowledgments}

\bibliography{note}

\providecommand{\noopsort}[1]{}\providecommand{\singleletter}[1]{#1}%
\begin{thebibliography}{37}%
\makeatletter
\providecommand \@ifxundefined [1]{%
 \@ifx{#1\undefined}
}%
\providecommand \@ifnum [1]{%
 \ifnum #1\expandafter \@firstoftwo
 \else \expandafter \@secondoftwo
 \fi
}%
\providecommand \@ifx [1]{%
 \ifx #1\expandafter \@firstoftwo
 \else \expandafter \@secondoftwo
 \fi
}%
\providecommand \natexlab [1]{#1}%
\providecommand \enquote  [1]{``#1''}%
\providecommand \bibnamefont  [1]{#1}%
\providecommand \bibfnamefont [1]{#1}%
\providecommand \citenamefont [1]{#1}%
\providecommand \href@noop [0]{\@secondoftwo}%
\providecommand \href [0]{\begingroup \@sanitize@url \@href}%
\providecommand \@href[1]{\@@startlink{#1}\@@href}%
\providecommand \@@href[1]{\endgroup#1\@@endlink}%
\providecommand \@sanitize@url [0]{\catcode `\\12\catcode `\$12\catcode
  `\&12\catcode `\#12\catcode `\^12\catcode `\_12\catcode `\%12\relax}%
\providecommand \@@startlink[1]{}%
\providecommand \@@endlink[0]{}%
\providecommand \url  [0]{\begingroup\@sanitize@url \@url }%
\providecommand \@url [1]{\endgroup\@href {#1}{\urlprefix }}%
\providecommand \urlprefix  [0]{URL }%
\providecommand \Eprint [0]{\href }%
\providecommand \doibase [0]{http://dx.doi.org/}%
\providecommand \selectlanguage [0]{\@gobble}%
\providecommand \bibinfo  [0]{\@secondoftwo}%
\providecommand \bibfield  [0]{\@secondoftwo}%
\providecommand \translation [1]{[#1]}%
\providecommand \BibitemOpen [0]{}%
\providecommand \bibitemStop [0]{}%
\providecommand \bibitemNoStop [0]{.\EOS\space}%
\providecommand \EOS [0]{\spacefactor3000\relax}%
\providecommand \BibitemShut  [1]{\csname bibitem#1\endcsname}%
\let\auto@bib@innerbib\@empty
\bibitem [{\citenamefont {Solodukhin}(2011)}]{solodukhin2011entanglement}%
  \BibitemOpen
  \bibfield  {author} {\bibinfo {author} {\bibfnamefont {S.~N.}\ \bibnamefont
  {Solodukhin}},\ }\href@noop {} {\bibfield  {journal} {\bibinfo  {journal}
  {Living Rev. Rel}\ }\textbf {\bibinfo {volume} {14}},\ \bibinfo {pages}
  {1104} (\bibinfo {year} {2011})}\BibitemShut {NoStop}%
\bibitem [{\citenamefont {Vidal}\ \emph {et~al.}(2003)\citenamefont {Vidal},
  \citenamefont {Latorre}, \citenamefont {Rico},\ and\ \citenamefont
  {Kitaev}}]{vidal2003entanglement}%
  \BibitemOpen
  \bibfield  {author} {\bibinfo {author} {\bibfnamefont {G.}~\bibnamefont
  {Vidal}}, \bibinfo {author} {\bibfnamefont {J.~I.}\ \bibnamefont {Latorre}},
  \bibinfo {author} {\bibfnamefont {E.}~\bibnamefont {Rico}}, \ and\ \bibinfo
  {author} {\bibfnamefont {A.}~\bibnamefont {Kitaev}},\ }\href@noop {}
  {\bibfield  {journal} {\bibinfo  {journal} {Physical review letters}\
  }\textbf {\bibinfo {volume} {90}},\ \bibinfo {pages} {227902} (\bibinfo
  {year} {2003})}\BibitemShut {NoStop}%
\bibitem [{\citenamefont {Pollmann}\ \emph {et~al.}(2009)\citenamefont
  {Pollmann}, \citenamefont {Mukerjee}, \citenamefont {Turner},\ and\
  \citenamefont {Moore}}]{pollmann2009theory}%
  \BibitemOpen
  \bibfield  {author} {\bibinfo {author} {\bibfnamefont {F.}~\bibnamefont
  {Pollmann}}, \bibinfo {author} {\bibfnamefont {S.}~\bibnamefont {Mukerjee}},
  \bibinfo {author} {\bibfnamefont {A.~M.}\ \bibnamefont {Turner}}, \ and\
  \bibinfo {author} {\bibfnamefont {J.~E.}\ \bibnamefont {Moore}},\ }\href@noop
  {} {\bibfield  {journal} {\bibinfo  {journal} {Physical review letters}\
  }\textbf {\bibinfo {volume} {102}},\ \bibinfo {pages} {255701} (\bibinfo
  {year} {2009})}\BibitemShut {NoStop}%
\bibitem [{\citenamefont {Kitaev}\ and\ \citenamefont
  {Preskill}(2006)}]{kitaev2006topological}%
  \BibitemOpen
  \bibfield  {author} {\bibinfo {author} {\bibfnamefont {A.}~\bibnamefont
  {Kitaev}}\ and\ \bibinfo {author} {\bibfnamefont {J.}~\bibnamefont
  {Preskill}},\ }\href@noop {} {\bibfield  {journal} {\bibinfo  {journal}
  {Physical review letters}\ }\textbf {\bibinfo {volume} {96}},\ \bibinfo
  {pages} {110404} (\bibinfo {year} {2006})}\BibitemShut {NoStop}%
\bibitem [{\citenamefont {Berkovits}(2012)}]{berkovits2012entanglement}%
  \BibitemOpen
  \bibfield  {author} {\bibinfo {author} {\bibfnamefont {R.}~\bibnamefont
  {Berkovits}},\ }\href@noop {} {\bibfield  {journal} {\bibinfo  {journal}
  {Physical review letters}\ }\textbf {\bibinfo {volume} {108}},\ \bibinfo
  {pages} {176803} (\bibinfo {year} {2012})}\BibitemShut {NoStop}%
\bibitem [{\citenamefont {Li}\ \emph {et~al.}(2016)\citenamefont {Li},
  \citenamefont {Pixley}, \citenamefont {Deng}, \citenamefont {Ganeshan},\ and\
  \citenamefont {Sarma}}]{li2016quantum}%
  \BibitemOpen
  \bibfield  {author} {\bibinfo {author} {\bibfnamefont {X.}~\bibnamefont
  {Li}}, \bibinfo {author} {\bibfnamefont {J.}~\bibnamefont {Pixley}}, \bibinfo
  {author} {\bibfnamefont {D.-L.}\ \bibnamefont {Deng}}, \bibinfo {author}
  {\bibfnamefont {S.}~\bibnamefont {Ganeshan}}, \ and\ \bibinfo {author}
  {\bibfnamefont {S.~D.}\ \bibnamefont {Sarma}},\ }\href@noop {} {\bibfield
  {journal} {\bibinfo  {journal} {Physical Review B}\ }\textbf {\bibinfo
  {volume} {93}},\ \bibinfo {pages} {184204} (\bibinfo {year}
  {2016})}\BibitemShut {NoStop}%
\bibitem [{\citenamefont {Yu}\ \emph {et~al.}(2016)\citenamefont {Yu},
  \citenamefont {Luitz},\ and\ \citenamefont {Clark}}]{yu2016bimodal}%
  \BibitemOpen
  \bibfield  {author} {\bibinfo {author} {\bibfnamefont {X.}~\bibnamefont
  {Yu}}, \bibinfo {author} {\bibfnamefont {D.~J.}\ \bibnamefont {Luitz}}, \
  and\ \bibinfo {author} {\bibfnamefont {B.~K.}\ \bibnamefont {Clark}},\
  }\href@noop {} {\bibfield  {journal} {\bibinfo  {journal} {arXiv preprint
  arXiv:1606.01260}\ } (\bibinfo {year} {2016})}\BibitemShut {NoStop}%
\bibitem [{\citenamefont {Bhattacharya}\ \emph {et~al.}(2013)\citenamefont
  {Bhattacharya}, \citenamefont {Nozaki}, \citenamefont {Takayanagi},\ and\
  \citenamefont {Ugajin}}]{bhattacharya2013thermodynamical}%
  \BibitemOpen
  \bibfield  {author} {\bibinfo {author} {\bibfnamefont {J.}~\bibnamefont
  {Bhattacharya}}, \bibinfo {author} {\bibfnamefont {M.}~\bibnamefont
  {Nozaki}}, \bibinfo {author} {\bibfnamefont {T.}~\bibnamefont {Takayanagi}},
  \ and\ \bibinfo {author} {\bibfnamefont {T.}~\bibnamefont {Ugajin}},\
  }\href@noop {} {\bibfield  {journal} {\bibinfo  {journal} {Physical review
  letters}\ }\textbf {\bibinfo {volume} {110}},\ \bibinfo {pages} {091602}
  (\bibinfo {year} {2013})}\BibitemShut {NoStop}%
\bibitem [{\citenamefont {P{\'a}lmai}(2014)}]{palmai2014excited}%
  \BibitemOpen
  \bibfield  {author} {\bibinfo {author} {\bibfnamefont {T.}~\bibnamefont
  {P{\'a}lmai}},\ }\href@noop {} {\bibfield  {journal} {\bibinfo  {journal}
  {Physical Review B}\ }\textbf {\bibinfo {volume} {90}},\ \bibinfo {pages}
  {161404} (\bibinfo {year} {2014})}\BibitemShut {NoStop}%
\bibitem [{\citenamefont {Lai}\ and\ \citenamefont
  {Yang}(2015)}]{lai2015entanglement}%
  \BibitemOpen
  \bibfield  {author} {\bibinfo {author} {\bibfnamefont {H.-H.}\ \bibnamefont
  {Lai}}\ and\ \bibinfo {author} {\bibfnamefont {K.}~\bibnamefont {Yang}},\
  }\href@noop {} {\bibfield  {journal} {\bibinfo  {journal} {Physical Review
  B}\ }\textbf {\bibinfo {volume} {91}},\ \bibinfo {pages} {081110} (\bibinfo
  {year} {2015})}\BibitemShut {NoStop}%
\bibitem [{\citenamefont {Hastings}(2007)}]{hastings2007area}%
  \BibitemOpen
  \bibfield  {author} {\bibinfo {author} {\bibfnamefont {M.~B.}\ \bibnamefont
  {Hastings}},\ }\href@noop {} {\bibfield  {journal} {\bibinfo  {journal}
  {Journal of Statistical Mechanics: Theory and Experiment}\ }\textbf {\bibinfo
  {volume} {2007}},\ \bibinfo {pages} {P08024} (\bibinfo {year}
  {2007})}\BibitemShut {NoStop}%
\bibitem [{\citenamefont {White}(1992)}]{white1992density}%
  \BibitemOpen
  \bibfield  {author} {\bibinfo {author} {\bibfnamefont {S.~R.}\ \bibnamefont
  {White}},\ }\href@noop {} {\bibfield  {journal} {\bibinfo  {journal}
  {Physical Review Letters}\ }\textbf {\bibinfo {volume} {69}},\ \bibinfo
  {pages} {2863} (\bibinfo {year} {1992})}\BibitemShut {NoStop}%
\bibitem [{\citenamefont {Schollw{\"o}ck}(2011)}]{schollwock2011density}%
  \BibitemOpen
  \bibfield  {author} {\bibinfo {author} {\bibfnamefont {U.}~\bibnamefont
  {Schollw{\"o}ck}},\ }\href@noop {} {\bibfield  {journal} {\bibinfo  {journal}
  {Annals of Physics}\ }\textbf {\bibinfo {volume} {326}},\ \bibinfo {pages}
  {96} (\bibinfo {year} {2011})}\BibitemShut {NoStop}%
\bibitem [{\citenamefont {Srednicki}(1993)}]{srednicki1993entropy}%
  \BibitemOpen
  \bibfield  {author} {\bibinfo {author} {\bibfnamefont {M.}~\bibnamefont
  {Srednicki}},\ }\href@noop {} {\bibfield  {journal} {\bibinfo  {journal}
  {Physical Review Letters}\ }\textbf {\bibinfo {volume} {71}},\ \bibinfo
  {pages} {666} (\bibinfo {year} {1993})}\BibitemShut {NoStop}%
\bibitem [{\citenamefont {Eisert}\ \emph {et~al.}(2010)\citenamefont {Eisert},
  \citenamefont {Cramer},\ and\ \citenamefont {Plenio}}]{eisert2010colloquium}%
  \BibitemOpen
  \bibfield  {author} {\bibinfo {author} {\bibfnamefont {J.}~\bibnamefont
  {Eisert}}, \bibinfo {author} {\bibfnamefont {M.}~\bibnamefont {Cramer}}, \
  and\ \bibinfo {author} {\bibfnamefont {M.~B.}\ \bibnamefont {Plenio}},\
  }\href@noop {} {\bibfield  {journal} {\bibinfo  {journal} {Reviews of Modern
  Physics}\ }\textbf {\bibinfo {volume} {82}},\ \bibinfo {pages} {277}
  (\bibinfo {year} {2010})}\BibitemShut {NoStop}%
\bibitem [{\citenamefont {Calabrese}\ and\ \citenamefont
  {Cardy}(2009)}]{calabrese2009entanglement}%
  \BibitemOpen
  \bibfield  {author} {\bibinfo {author} {\bibfnamefont {P.}~\bibnamefont
  {Calabrese}}\ and\ \bibinfo {author} {\bibfnamefont {J.}~\bibnamefont
  {Cardy}},\ }\href@noop {} {\bibfield  {journal} {\bibinfo  {journal} {Journal
  of Physics A: Mathematical and Theoretical}\ }\textbf {\bibinfo {volume}
  {42}},\ \bibinfo {pages} {504005} (\bibinfo {year} {2009})}\BibitemShut
  {NoStop}%
\bibitem [{\citenamefont {Gioev}\ and\ \citenamefont
  {Klich}(2006)}]{gioev2006entanglement}%
  \BibitemOpen
  \bibfield  {author} {\bibinfo {author} {\bibfnamefont {D.}~\bibnamefont
  {Gioev}}\ and\ \bibinfo {author} {\bibfnamefont {I.}~\bibnamefont {Klich}},\
  }\href@noop {} {\bibfield  {journal} {\bibinfo  {journal} {Physical review
  letters}\ }\textbf {\bibinfo {volume} {96}},\ \bibinfo {pages} {100503}
  (\bibinfo {year} {2006})}\BibitemShut {NoStop}%
\bibitem [{\citenamefont {Refael}\ and\ \citenamefont
  {Moore}(2004)}]{refael2004entanglement}%
  \BibitemOpen
  \bibfield  {author} {\bibinfo {author} {\bibfnamefont {G.}~\bibnamefont
  {Refael}}\ and\ \bibinfo {author} {\bibfnamefont {J.~E.}\ \bibnamefont
  {Moore}},\ }\href@noop {} {\bibfield  {journal} {\bibinfo  {journal}
  {Physical review letters}\ }\textbf {\bibinfo {volume} {93}},\ \bibinfo
  {pages} {260602} (\bibinfo {year} {2004})}\BibitemShut {NoStop}%
\bibitem [{\citenamefont {Lin}\ \emph {et~al.}(2007)\citenamefont {Lin},
  \citenamefont {Igl{\'o}i},\ and\ \citenamefont
  {Rieger}}]{lin2007entanglement}%
  \BibitemOpen
  \bibfield  {author} {\bibinfo {author} {\bibfnamefont {Y.-C.}\ \bibnamefont
  {Lin}}, \bibinfo {author} {\bibfnamefont {F.}~\bibnamefont {Igl{\'o}i}}, \
  and\ \bibinfo {author} {\bibfnamefont {H.}~\bibnamefont {Rieger}},\
  }\href@noop {} {\bibfield  {journal} {\bibinfo  {journal} {Physical review
  letters}\ }\textbf {\bibinfo {volume} {99}},\ \bibinfo {pages} {147202}
  (\bibinfo {year} {2007})}\BibitemShut {NoStop}%
\bibitem [{\citenamefont {Calabrese}\ \emph {et~al.}(2010)\citenamefont
  {Calabrese}, \citenamefont {Cardy},\ and\ \citenamefont
  {Peschel}}]{calabrese2010corrections}%
  \BibitemOpen
  \bibfield  {author} {\bibinfo {author} {\bibfnamefont {P.}~\bibnamefont
  {Calabrese}}, \bibinfo {author} {\bibfnamefont {J.}~\bibnamefont {Cardy}}, \
  and\ \bibinfo {author} {\bibfnamefont {I.}~\bibnamefont {Peschel}},\
  }\href@noop {} {\bibfield  {journal} {\bibinfo  {journal} {Journal of
  Statistical Mechanics: Theory and Experiment}\ }\textbf {\bibinfo {volume}
  {2010}},\ \bibinfo {pages} {P09003} (\bibinfo {year} {2010})}\BibitemShut
  {NoStop}%
\bibitem [{\citenamefont {Rodney}\ \emph {et~al.}(2013)\citenamefont {Rodney},
  \citenamefont {Song}, \citenamefont {Lee}, \citenamefont {Le~Hur},\ and\
  \citenamefont {S{\o}rensen}}]{rodney2013scaling}%
  \BibitemOpen
  \bibfield  {author} {\bibinfo {author} {\bibfnamefont {M.}~\bibnamefont
  {Rodney}}, \bibinfo {author} {\bibfnamefont {H.~F.}\ \bibnamefont {Song}},
  \bibinfo {author} {\bibfnamefont {S.-S.}\ \bibnamefont {Lee}}, \bibinfo
  {author} {\bibfnamefont {K.}~\bibnamefont {Le~Hur}}, \ and\ \bibinfo {author}
  {\bibfnamefont {E.~S.}\ \bibnamefont {S{\o}rensen}},\ }\href@noop {}
  {\bibfield  {journal} {\bibinfo  {journal} {Physical Review B}\ }\textbf
  {\bibinfo {volume} {87}},\ \bibinfo {pages} {115132} (\bibinfo {year}
  {2013})}\BibitemShut {NoStop}%
\bibitem [{\citenamefont {Peschel}(2003)}]{peschel2003calculation}%
  \BibitemOpen
  \bibfield  {author} {\bibinfo {author} {\bibfnamefont {I.}~\bibnamefont
  {Peschel}},\ }\href@noop {} {\bibfield  {journal} {\bibinfo  {journal}
  {Journal of Physics A: Mathematical and General}\ }\textbf {\bibinfo {volume}
  {36}},\ \bibinfo {pages} {L205} (\bibinfo {year} {2003})}\BibitemShut
  {NoStop}%
\bibitem [{\citenamefont {Cheong}\ and\ \citenamefont
  {Henley}(2004)}]{cheong2004many}%
  \BibitemOpen
  \bibfield  {author} {\bibinfo {author} {\bibfnamefont {S.-A.}\ \bibnamefont
  {Cheong}}\ and\ \bibinfo {author} {\bibfnamefont {C.~L.}\ \bibnamefont
  {Henley}},\ }\href@noop {} {\bibfield  {journal} {\bibinfo  {journal}
  {Physical Review B}\ }\textbf {\bibinfo {volume} {69}},\ \bibinfo {pages}
  {075111} (\bibinfo {year} {2004})}\BibitemShut {NoStop}%
\bibitem [{\citenamefont {Kitaev}(2001)}]{kitaev2001unpaired}%
  \BibitemOpen
  \bibfield  {author} {\bibinfo {author} {\bibfnamefont {A.~Y.}\ \bibnamefont
  {Kitaev}},\ }\href@noop {} {\bibfield  {journal} {\bibinfo  {journal}
  {Physics-Uspekhi}\ }\textbf {\bibinfo {volume} {44}},\ \bibinfo {pages} {131}
  (\bibinfo {year} {2001})}\BibitemShut {NoStop}%
\bibitem [{\citenamefont {Eisler}\ and\ \citenamefont
  {Peschel}(2010)}]{eisler2010entanglement}%
  \BibitemOpen
  \bibfield  {author} {\bibinfo {author} {\bibfnamefont {V.}~\bibnamefont
  {Eisler}}\ and\ \bibinfo {author} {\bibfnamefont {I.}~\bibnamefont
  {Peschel}},\ }\href@noop {} {\bibfield  {journal} {\bibinfo  {journal}
  {Annalen der Physik}\ }\textbf {\bibinfo {volume} {522}},\ \bibinfo {pages}
  {679} (\bibinfo {year} {2010})}\BibitemShut {NoStop}%
\bibitem [{\citenamefont {S{\"u}sstrunk}\ and\ \citenamefont
  {Ivanov}(2013)}]{susstrunk2013free}%
  \BibitemOpen
  \bibfield  {author} {\bibinfo {author} {\bibfnamefont {R.}~\bibnamefont
  {S{\"u}sstrunk}}\ and\ \bibinfo {author} {\bibfnamefont {D.~A.}\ \bibnamefont
  {Ivanov}},\ }\href@noop {} {\bibfield  {journal} {\bibinfo  {journal} {EPL
  (Europhysics Letters)}\ }\textbf {\bibinfo {volume} {100}},\ \bibinfo {pages}
  {60009} (\bibinfo {year} {2013})}\BibitemShut {NoStop}%
\bibitem [{\citenamefont {Song}\ \emph {et~al.}(2010)\citenamefont {Song},
  \citenamefont {Rachel},\ and\ \citenamefont {Le~Hur}}]{song2010general}%
  \BibitemOpen
  \bibfield  {author} {\bibinfo {author} {\bibfnamefont {H.~F.}\ \bibnamefont
  {Song}}, \bibinfo {author} {\bibfnamefont {S.}~\bibnamefont {Rachel}}, \ and\
  \bibinfo {author} {\bibfnamefont {K.}~\bibnamefont {Le~Hur}},\ }\href@noop {}
  {\bibfield  {journal} {\bibinfo  {journal} {Physical Review B}\ }\textbf
  {\bibinfo {volume} {82}},\ \bibinfo {pages} {012405} (\bibinfo {year}
  {2010})}\BibitemShut {NoStop}%
\bibitem [{\citenamefont {Eisler}\ \emph {et~al.}(2015)\citenamefont {Eisler},
  \citenamefont {Chung},\ and\ \citenamefont
  {Peschel}}]{eisler2015entanglement}%
  \BibitemOpen
  \bibfield  {author} {\bibinfo {author} {\bibfnamefont {V.}~\bibnamefont
  {Eisler}}, \bibinfo {author} {\bibfnamefont {M.-C.}\ \bibnamefont {Chung}}, \
  and\ \bibinfo {author} {\bibfnamefont {I.}~\bibnamefont {Peschel}},\
  }\href@noop {} {\bibfield  {journal} {\bibinfo  {journal} {Journal of
  Statistical Mechanics: Theory and Experiment}\ }\textbf {\bibinfo {volume}
  {2015}},\ \bibinfo {pages} {P07011} (\bibinfo {year} {2015})}\BibitemShut
  {NoStop}%
\bibitem [{\citenamefont {Peschel}(2005)}]{peschel2005entanglement}%
  \BibitemOpen
  \bibfield  {author} {\bibinfo {author} {\bibfnamefont {I.}~\bibnamefont
  {Peschel}},\ }\href@noop {} {\bibfield  {journal} {\bibinfo  {journal}
  {Journal of Physics A: Mathematical and General}\ }\textbf {\bibinfo {volume}
  {38}},\ \bibinfo {pages} {4327} (\bibinfo {year} {2005})}\BibitemShut
  {NoStop}%
\bibitem [{\citenamefont {Eggert}\ and\ \citenamefont
  {Affleck}(1992)}]{eggert1992magnetic}%
  \BibitemOpen
  \bibfield  {author} {\bibinfo {author} {\bibfnamefont {S.}~\bibnamefont
  {Eggert}}\ and\ \bibinfo {author} {\bibfnamefont {I.}~\bibnamefont
  {Affleck}},\ }\href@noop {} {\bibfield  {journal} {\bibinfo  {journal}
  {Physical Review B}\ }\textbf {\bibinfo {volume} {46}},\ \bibinfo {pages}
  {10866} (\bibinfo {year} {1992})}\BibitemShut {NoStop}%
\bibitem [{\citenamefont {Zhao}\ \emph {et~al.}(2006)\citenamefont {Zhao},
  \citenamefont {Peschel},\ and\ \citenamefont {Wang}}]{zhao2006critical}%
  \BibitemOpen
  \bibfield  {author} {\bibinfo {author} {\bibfnamefont {J.}~\bibnamefont
  {Zhao}}, \bibinfo {author} {\bibfnamefont {I.}~\bibnamefont {Peschel}}, \
  and\ \bibinfo {author} {\bibfnamefont {X.}~\bibnamefont {Wang}},\ }\href@noop
  {} {\bibfield  {journal} {\bibinfo  {journal} {Physical Review B}\ }\textbf
  {\bibinfo {volume} {73}},\ \bibinfo {pages} {024417} (\bibinfo {year}
  {2006})}\BibitemShut {NoStop}%
\bibitem [{\citenamefont {S{\o}rensen}\ \emph {et~al.}(2007)\citenamefont
  {S{\o}rensen}, \citenamefont {Chang}, \citenamefont {Laflorencie},\ and\
  \citenamefont {Affleck}}]{sorensen2007quantum}%
  \BibitemOpen
  \bibfield  {author} {\bibinfo {author} {\bibfnamefont {E.~S.}\ \bibnamefont
  {S{\o}rensen}}, \bibinfo {author} {\bibfnamefont {M.-S.}\ \bibnamefont
  {Chang}}, \bibinfo {author} {\bibfnamefont {N.}~\bibnamefont {Laflorencie}},
  \ and\ \bibinfo {author} {\bibfnamefont {I.}~\bibnamefont {Affleck}},\
  }\href@noop {} {\bibfield  {journal} {\bibinfo  {journal} {Journal of
  Statistical Mechanics: Theory and Experiment}\ }\textbf {\bibinfo {volume}
  {2007}},\ \bibinfo {pages} {P08003} (\bibinfo {year} {2007})}\BibitemShut
  {NoStop}%
\bibitem [{\citenamefont {Igl{\'o}i}\ \emph {et~al.}(2009)\citenamefont
  {Igl{\'o}i}, \citenamefont {Szatm{\'a}ri},\ and\ \citenamefont
  {Lin}}]{igloi2009entanglement}%
  \BibitemOpen
  \bibfield  {author} {\bibinfo {author} {\bibfnamefont {F.}~\bibnamefont
  {Igl{\'o}i}}, \bibinfo {author} {\bibfnamefont {Z.}~\bibnamefont
  {Szatm{\'a}ri}}, \ and\ \bibinfo {author} {\bibfnamefont {Y.-C.}\
  \bibnamefont {Lin}},\ }\href@noop {} {\bibfield  {journal} {\bibinfo
  {journal} {Physical Review B}\ }\textbf {\bibinfo {volume} {80}},\ \bibinfo
  {pages} {024405} (\bibinfo {year} {2009})}\BibitemShut {NoStop}%
\bibitem [{\citenamefont {Saleur}\ \emph {et~al.}(2013)\citenamefont {Saleur},
  \citenamefont {Schmitteckert},\ and\ \citenamefont
  {Vasseur}}]{saleur2013entanglement}%
  \BibitemOpen
  \bibfield  {author} {\bibinfo {author} {\bibfnamefont {H.}~\bibnamefont
  {Saleur}}, \bibinfo {author} {\bibfnamefont {P.}~\bibnamefont
  {Schmitteckert}}, \ and\ \bibinfo {author} {\bibfnamefont {R.}~\bibnamefont
  {Vasseur}},\ }\href@noop {} {\bibfield  {journal} {\bibinfo  {journal}
  {Physical Review B}\ }\textbf {\bibinfo {volume} {88}},\ \bibinfo {pages}
  {085413} (\bibinfo {year} {2013})}\BibitemShut {NoStop}%
\bibitem [{\citenamefont {Ossipov}(2014)}]{ossipov2014entanglement}%
  \BibitemOpen
  \bibfield  {author} {\bibinfo {author} {\bibfnamefont {A.}~\bibnamefont
  {Ossipov}},\ }\href@noop {} {\bibfield  {journal} {\bibinfo  {journal}
  {Physical review letters}\ }\textbf {\bibinfo {volume} {113}},\ \bibinfo
  {pages} {130402} (\bibinfo {year} {2014})}\BibitemShut {NoStop}%
\bibitem [{\citenamefont {Potter}\ \emph {et~al.}(2015)\citenamefont {Potter},
  \citenamefont {Vasseur},\ and\ \citenamefont
  {Parameswaran}}]{potter2015universal}%
  \BibitemOpen
  \bibfield  {author} {\bibinfo {author} {\bibfnamefont {A.~C.}\ \bibnamefont
  {Potter}}, \bibinfo {author} {\bibfnamefont {R.}~\bibnamefont {Vasseur}}, \
  and\ \bibinfo {author} {\bibfnamefont {S.}~\bibnamefont {Parameswaran}},\
  }\href@noop {} {\bibfield  {journal} {\bibinfo  {journal} {Physical Review
  X}\ }\textbf {\bibinfo {volume} {5}},\ \bibinfo {pages} {031033} (\bibinfo
  {year} {2015})}\BibitemShut {NoStop}%
\bibitem [{\citenamefont {McCann}\ and\ \citenamefont
  {Koshino}(2013)}]{mccann2013electronic}%
  \BibitemOpen
  \bibfield  {author} {\bibinfo {author} {\bibfnamefont {E.}~\bibnamefont
  {McCann}}\ and\ \bibinfo {author} {\bibfnamefont {M.}~\bibnamefont
  {Koshino}},\ }\href@noop {} {\bibfield  {journal} {\bibinfo  {journal}
  {Reports on Progress in Physics}\ }\textbf {\bibinfo {volume} {76}},\
  \bibinfo {pages} {056503} (\bibinfo {year} {2013})}\BibitemShut {NoStop}%
\end{thebibliography}%

\end{document}